\documentclass[letterpaper]{article}
\usepackage{iccc}
\usepackage{times}
\usepackage{helvet}
\usepackage{courier}
\usepackage{multirow} 
\usepackage{array} 
\usepackage{tabularx} 
\usepackage{makecell}
\usepackage{booktabs} 
\usepackage{float}
\usepackage{color}
\usepackage{pdfpages}
\usepackage{amsmath}
\usepackage[hyphens]{url}
\usepackage{algpseudocode}
\newcommand{\leftcolumnfootnotes}{%
  \makeatletter
  \renewcommand\@makefntext[1]{%
    \parindent 1em%
    \noindent\hb@xt@1.8em{\hss\@makefnmark}##1}%
  \interfootnotelinepenalty=10000
  \renewcommand\footnoterule{%
    \kern-3\p@
    \hrule width 2in height 0.4pt
    \kern 2.6\p}%
  \makeatother
}

\newcommand{\resetnormalfootnotes}{%
  \makeatletter
  \interfootnotelinepenalty=0
  \renewcommand\footnoterule{%
    \kern-3\p
    \hrule width \columnwidth height 0.4pt
    \kern 2.6\p}%
  \makeatother
}
\title{Do Conversational Interfaces Limit Creativity? Exploring Visual Graph Systems for Creative Writing}
\author{Abhinav Sood\\
University of Sydney\\
Sydney, Australia\\
abhinav.sood@sydney.edu.au\\
\And
Maria Teresa Llano \\
University of Sussex\\
Brighton, UK\\
Teresa.Llano@sussex.ac.uk\\
\And
Jon McCormack \\
Sensilab, Monash University\\
Caulfield East, Australia\\
Jon.McCormack@monash.edu\\
}

\begin{document} 
\maketitle
\begin{abstract}
\begin{quote}
  We present a graphical, node-based system through which users can visually chain generative AI models for creative tasks. Research in chaining LLMs has found that while chaining provides transparency, controllability and guardrails to approach certain tasks, chaining with pre-defined LLM steps prevents free exploration. Using cognitive processes from creativity research as a basis, we create a system that addresses the inherent constraints of chat-based AI interactions. Specifically, our system aims to overcome the limiting linear structure that inhibits creative exploration and ideation. Further, our node-based approach enables the creation of reusable, shareable templates that can address different creative tasks. In a small-scale user study, we find that our graph-based system supports ideation and allows some users to better visualise and think through their writing process when compared to a similar conversational interface. We further discuss the weaknesses and limitations of our system, noting the benefits to creativity that user interfaces with higher complexity can provide for users who can effectively use them.
\end{quote}
\end{abstract}

\section{Introduction}
The creative writing process often involves relying on ideation and open-ended exploration to complete tasks that can then be used to enhance written material. For example, in the fantasy and sci-fi genres, world creation, character description and characters' interactions with the world are crucial to writing good stories \cite{card2001write}. Large Language Models (LLMs) have emerged as generative AI tools that can support the natural language components of such tasks. For many users of LLMs, interactions often occur through linear text-based chat interfaces that pose significant limitations. Their inherent linear structure prevents exploration through branching, and well-engineered prompts are required to guide LLMs, preventing the effective use of these tools by everyday users \cite{johnny}. These challenges only increase for more complicated writing tasks. Here, novel interfaces like Metaphorian \cite{metaphorian} and AngleKindling \cite{AngleKindling} have been developed for tasks like finding metaphors for scientific writing and exploring journalistic angles for press releases. However, as these tools are tailored to specific writing domains, they do not provide a generalised platform for working on open-ended writing problems.

To address this gap, we present a graphical, node-based system that allows users to chain generative AI models and prompts. Once these chains are set up, they can be saved into shareable templates that end-users can then easily use for more complicated creative writing tasks without having to engineer prompts. We then investigate the properties of this system in a creative writing context. Our key contributions are as follows:
\begin{enumerate}
	\item We develop a general-purpose, openly available, graph-based system that uses chaining with relaxed restrictions. This system can be used for different writing tasks. 
     \footnote{The code for the system, along with additional documentation, is available at - {\color{blue} \url{https://github.com/abhinavsood2002/Chaining-GenAI-for-Ideation}}}
    \item We conduct a user study to assess the effectiveness of our system for planning related creative writing tasks that involve ideation and open-ended exploration. We find that our graph-based system illustrates how different users benefit from different interfaces for ideation.
    \item Our user study provides us with insights into how such interfaces affect users' writing processes. We discuss how these insights can help develop future interfaces in this area.
\end{enumerate}

\begin{figure*}[ht!]
	\centering
	\includegraphics[width=\textwidth]{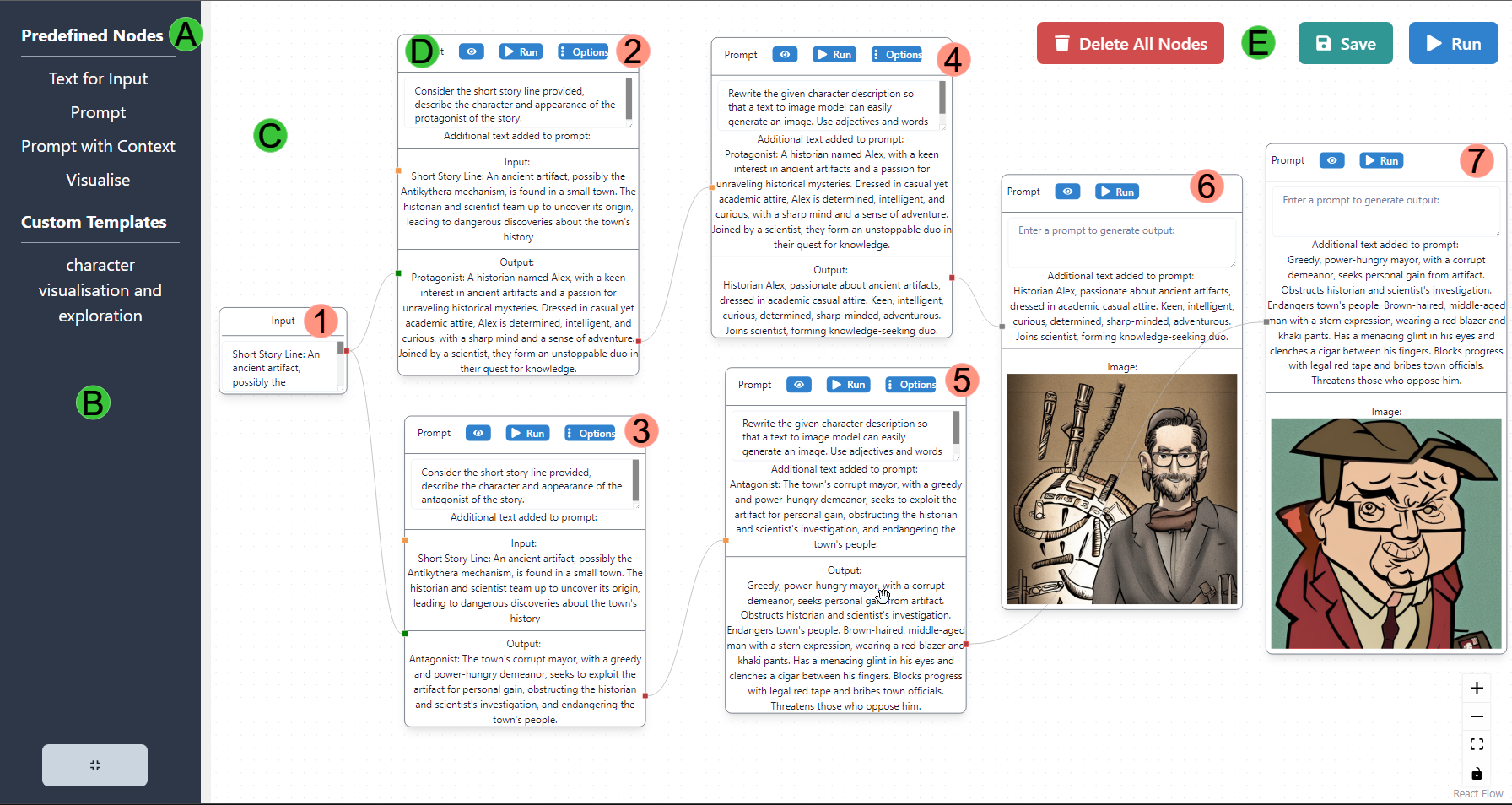}
	\caption[The User Interface alongside the demonstration of an example use-case where the developed system can be used to visualise and explore characters.]%
 {The User Interface alongside the demonstration of an example use-case where the developed system can be used to visualise and explore characters. \par  \textbf{User Interface (Green)} - (A) The 4 Predefined Node options. (B) Custom Node section where templates can be dropped in to make them available in the user interface. (C) The workspace - nodes from the sidebar can be dragged and dropped into the workspace and subsequently edited. (D) The node header contains buttons that activate additional functionality - like running individual nodes, viewing node information in a user-friendly Modal, and changing node-specific parameters like temperature for generation. (E) The save button downloads the current workspace as a template as a JSON file, and the run button runs all nodes in topological order.  \\ \textbf{Character Visualisation and Exploration Use Case (Orange)} - (1) contains a short storyline. (2) and (3) use the short storyline to create a protagonist and antagonist character description. (4) and (5) are used to prepare the descriptions for visualisation by converting them into a form that is more friendly for text-to-image prompting. (6) and (7) display visualisations for the characters.}
	\label{fig:example}
\end{figure*}
\section{Background}
\subsection{Creativity and Ideation}
   In creativity research, three main cognitive processes have been essential \cite{lee2013cognitive} to understanding the psychological basis of creativity. These are divergent thinking, convergent thinking and associative processing. Divergent thinking is a cognitive process that generates a broad array of unconventional ideas by exploring various perspectives that might not be rational. In contrast, convergent thinking focuses on finding the best answer to a well-defined problem through rational reasoning. Associative processing involves connecting and linking concepts based on shared attributes. Together, these processes contribute to effective problem-solving and innovation.
    
These general cognitive processes emerge as creative behaviour in different forms. In creative writing, authors in fantasy often employ divergent thinking to construct diverse worlds. Strategies in ideation also relate to these cognitive processes. For example, divergent and convergent thinking play a role in ideation techniques like brainstorming \cite{osborn1953applied} and mind maps \cite{buzan1996mind}.

\subsection{Text and Image Generation}
The introduction of the transformer architecture \cite{trasformers} has brought rapid advances in language modelling. The architecture's self-attention mechanism allowed for efficient parallelisation while capturing long-range dependencies, making it feasible to train language models on massive datasets. This, in turn, has led to the development of Large Language Models (LLMs) like GPT-3 \cite{few-shot}. Inference with LLMs occurs through zero or few-shot prompts. While zero-shot prompts contain a natural language instruction, few-shot prompts additionally contain examples related to the instruction to provide the LLM with more context for approaching a problem. 
\par
 Advancements in language modelling have been accompanied by similar progress in the field of image generation. Denoising Diffusion Probabilistic Models (DDPMs) \cite{diffusionmodels} have been instrumental in unconditional image generation. DDPMs learn to generate images by modelling the gradual removal of noise from a noisy latent image. The conditioning of this noise removal process on text has led to the development of text-to-image models like Stable diffusion \cite{stablediffusion}. Both LLMs and text-to-image models hold significant potential for creative generation tasks. With their ability to generate human-like text, LLMs can be employed in storytelling, poetry generation, and scriptwriting \cite{kaddour2023challenges}. On the other hand, text-to-image can be utilised in visual art generation and story visualisation \cite{zhang2023texttoimage}.

\subsection{Human-AI Interaction and Creativity Support Tools}
The applications of generative AI models go beyond simple generation tasks. Human-AI interaction and co-creative applications are receiving increasing attention. Often, human-AI interaction involves a mixed-initiative system \cite{allen1999mixed} where both the machine and the system contribute towards a shared task. For creative tasks, constructing an adequate mixed-initiative system involves examining how generative AI tools can be incorporated into a user's creative workflow to facilitate collaboration between the user and AI. \citeauthor{karimi2020creative}(\citeyear{karimi2020creative}) provide an example of a system where an AI generates sketches that resemble a user's drawings but are from a different category. These generated sketches can then be used to re-interpret the original sketch in a new context, aiding their creative design process. Concurrent efforts have been made to augment and enhance the creative process of different types of writers. For example, Creativity Support Tools (CSTs) like Wordcraft \cite{wordcraft} can be used to perform text infilling after generating a passage, elaborating on a region of text, seeding the start of a story, and arbitrarily rewriting a section of text. Dramatron \cite{dramatron} provides another example of a tool that aids in co-writing screenplays and theatre scripts.

\subsection{Chaining LLMs and Node-Graph Interfaces}
CSTs that use LLMs often employ few-shot prompts (prompts with examples relating to the task) or chained prompts (sequences of LLM interactions where the output of the preceding chain interaction is passed to the following interaction) to produce valuable functionality. The concept of chaining prompts was explored in AI Chains \cite{aichains}, an interactive interface that allowed users to customise chains of prompts for given tasks. To the user, such interfaces present themselves as a node-graph interface, where fundamentally, nodes that can take input and output are used to organise a solution to a given problem. We use the terms node-graph interface and graph-based system interchangeably. Other interactive node-graph systems that use chaining include ChainForge \cite{chainforge}, PromptChainer \cite{PromptChainer}, and LangFlow\cite{langflow}. Neither of these systems focuses on how chaining can be adapted for creative exploration. ChainForge provides an interface to construct graphs to evaluate different language models against one another. PromptChainer aims to prototype software applications using LLMs, and Langflow caters to constructing AI applications. As such, these applications require knowledge of concepts in these domains and do not function as simple tools that support creative writing.

\section{Chaining for Creativity - A Graph-based System}
The lack of interfaces that use chaining for creative exploration prompted us to develop a graph-based interface for creativity. The two key design rationales for our system were:
\begin{enumerate}
    \item Allowing for \textbf{easy expansion and combination} of text.
    To ensure that our system could support creative tasks, we wanted our system to relate to the cognitive processes in creativity research. Specifically, we wanted to ensure that divergent and convergent thinking were supported by allowing for easy, LLM-based expansion, mutation, and combination of node outputs.
    \item Keeping interaction \textbf{simple}. Many graph-based systems that use chaining, like Flowise \cite{flowise} and LangFlow \cite{langflow}, require an understanding of the underlying machine-learning concepts. Since our end-user is a writer who might not be familiar with the technical details of machine learning, the nodes in our system must be simple to interact with.
    \item \textbf{Flexibility} in node definitions. Qualitative evidence from AI-Chains \cite{aichains} indicated that their interface was not ideal for free exploration, a key component of ideation. This was primarily because each interaction in the chain was a rigid, pre-defined operation. The authors suggested that structural constraints in each chaining step could be relaxed to encourage exploration for future work. 
    \end{enumerate}

Thus, we develop a graph-based system in which users primarily interact through simple and flexible nodes. Each node can be provided with multiple inputs (which are then concatenated) in the form of text. The input is processed into output based on a prompt that the user provides. The system is described in detail in the subsection that follows.

\subsection{System Description}
The interface developed in React.js is visible and described in Figure~1.
\subsubsection{Nodes}
Nodes are individual components of the system that can operate independently and perform a task of value to the user. Each node typically has connectors or handles along its boundaries, allowing it to be linked with other nodes. If a handle receives multiple inputs, the inputs are concatenated. For this system, we describe four fundamental predefined nodes:
\begin{itemize}
    \item \textbf{Text for Input:} This type of node serves as a placeholder for user input. Its output is merely a copy of the text in the node. It is used to organise input for a given chain.
	\item \textbf{Prompt:} This node allows the user to prompt an LLM for output. The prompt can be modified by adding additional text as input.
	\item \textbf{Prompt with Context:} This node is similar to the Prompt node but also includes an additional input that can be used to provide context that might not fit well in the prompt.
	\item \textbf{Visualise:} This node uses stable diffusion to generate images for a given prompt.
\end{itemize}

\subsubsection{Templates}
Once nodes are connected to form a graph, they can be saved into templates. Templates are saved as .JSON files. These files can be shared with other users of the system. This is possible due to the nature of graph-based interactions, where a node's output depends solely on the nodes from which it receives input. This would not be possible in a linear chat-based conversation, where one prompt changes all future prompts. As templates have well-defined inputs and outputs, multiple templates can be used in the same run to perform more complex tasks. During the execution of a template, the nodes are topologically sorted through breadth-first traversal. Then, each node is run in order. The algorithm used to perform this topological sorting is provided and described in the attached code repository.

\subsubsection{Additional features}
The system contains additional features to provide users with a higher degree of control over the interface and output. This includes the ability to set the variability of text (through temperature) for each node and the desired length of output from a node. Other visual components include a mini-map to navigate the interface for larger template setups and navigation tools to use the interface with a trackpad.

\subsubsection{Backend}
 We host our models on an NVIDIA RTX A6000 50GB GPU provided by SensiLab, Monash University. The decision to host our models locally was to ensure that any user studies conducted would not involve sending information to third-party services due to privacy concerns.

\section{User Study}
To test our system, we identified two planning-related creative writing tasks that our system could significantly impact. We identified \textbf{world-building} and \textbf{character description} as planning-related tasks of relevance. We conducted a small-scale user study with participants completing creative writing tasks to analyse and evaluate our system.

Eight participants were recruited through convenience sampling. The sampling method was selected due to the study design's requirement of approximately one and a half to two hours of participation time per participant. Given the absence of monetary compensation, recruiting participants without convenience sampling would not have been feasible. An explanatory statement was dispersed through creative writing-related social media groups and acquaintances. Those who were interested were provided with a consent form to confirm their participation in the study. Participants were provided with a 10-minute video \footnote{{\color{blue} \url{https://www.youtube.com/watch?v=ACVLsKSPN-0}}} guiding them through the interface and instructions for the study. We additionally construct a conversational interface as a baseline.

The study was conducted as follows: First, each participant received a link to an online interface along with a UserID. After logging in with their UserID, they were asked to complete a creative writing task using the graph-based and conversational system. Once they completed the tasks, they were given a questionnaire asking about their experiences with each system. Each participant contributed approximately 1:30-2 hours to complete the given tasks. The user study was approved through a Human Ethics Application made to the Monash University Human Research Ethics Committee.

\subsection{Baseline and Additional Changes}

The graph-based system had to be updated to conduct the user study virtually. All interaction with the system needs to occur in the user interface; this includes any material the participant writes down to complete a task. Thus, we added a notepad to the right-hand side of the interface where the user could draft the output for their tasks. 

We also constructed a conversational interface as a baseline. Previous user studies evaluating chaining have compared chaining LLMs to a sandbox(where LLMs provide answers to single instructions independently) environment \cite{aichains}. However, we believe that most user interactions with LLMs happen through conversational interfaces similar to ChatGPT\cite{OpenAI2024}. Furthermore, research on chatbot brainstorming has shown success against human baselines\cite{brainchatbot}. Thus, we compare our graph-based system against a comparable conversational interface (instead of a sandbox). Images demonstrating these user interfaces are available in the linked Code Repository.

The conversational interface builds on code available at \cite{onyebuchi_chatbot_2024}. We used the same LLM and system prompt to ensure that both the conversational and graph-based interfaces were comparable. The conversational interface was updated to generate images. Example queries for each task were also displayed. Preliminary testing with the conversational interface indicated that Mistral 7B \cite{mistral7b} Instruct v0.2 could not maintain context over multiple prompts when testing it on a creative writing task. Even though this LLM worked in our comprehensive example with the graph-based interface, it wasn't able to perform in the conversational interface. Thus, after testing out other LLMs, we chose OpenHermes-2.5-Mistral-7B\footnote{\color{blue}\url{https://huggingface.co/teknium/OpenHermes-2.5-Mistral-7B}} for its ability to maintain context in a conversation while still being a 7B model for compute restrictions. We use Stable Diffusion 2.1 \cite{stablediffusion} to generate images. 

\subsection{Study Design} 
\begin{figure*}[ht]
    \centering
    \includegraphics[width=\textwidth]{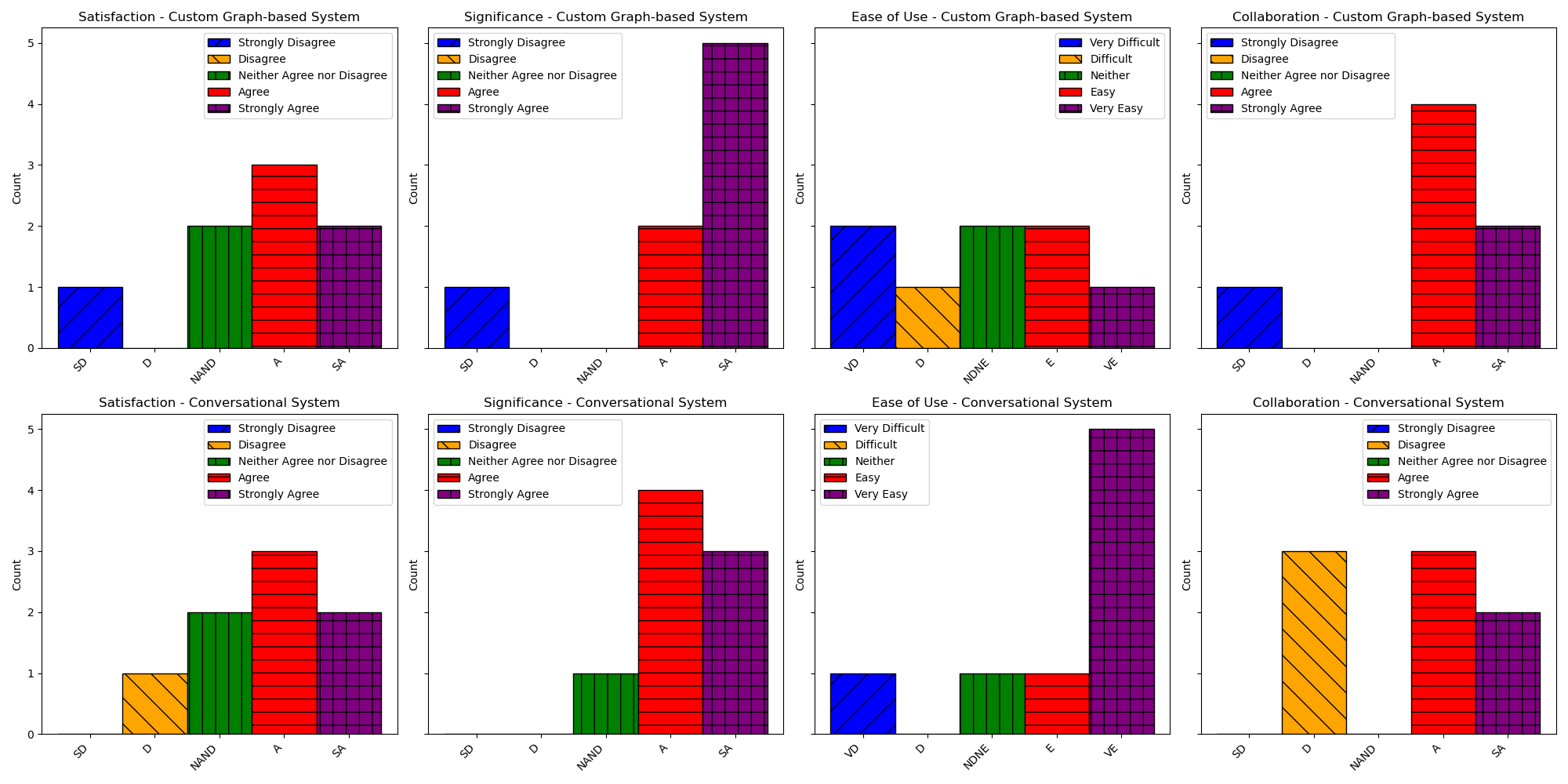}
    \caption{User Study Questionnaire - 5 point questions results.}
    \label{fig:results}
\end{figure*}

As our sample size was small, we wanted to minimise the influence of any confounding variables in our study. We identified two confounding variables:
\begin{enumerate}
    \item The order in which the systems were displayed could affect proficiency in interacting with the subsequent system.
    \item If the exact same creative writing task was used for both systems, a participant could use information from the previous task in the subsequent task.
\end{enumerate}
To address the first confounding variable, half of the participants were given the graph-based system first, while the other half were given the conversational system first. Participants were given variants of the same creative writing task for each system to address the second confounding variable. Two participants tested the system for each combination of order (graph-based or conversational first), creative writing task (world-building or character description) and variant.
\noindent
The specific creative writing tasks were:
\begin{itemize}
    \item \textbf{Character Description}
    \begin{itemize}
        \item Variant 1: Write a brief character description for a protagonist in a short story. The protagonist is Camilla, a librarian whose primary goal is to solve the mystery of her brother’s disappearance. She believes the disappearance is connected to a larger conspiracy.
        \item Variant 2:  Write a brief character description for an antagonist in a short story. The antagonist is General Kal, who ascended to power through a coup. He clamps down on any form of dissent, using all powers available to him to crush any opposition.
    \end{itemize}
    \item \textbf{World Building}
    \begin{itemize}
        \item Variant 1: Write a brief description of a world-setting for a community in a lush rainforest environment. Focus on the practices they have developed to sustain themselves.
        \item Variant 2: Write a brief description of a world-setting for a community in a harsh arctic environment. Detail their survival strategies in the extreme cold.
    \end{itemize}
\end{itemize}
\noindent
Setting up a series of nodes in the graph-based system might require a lot of experimentation. This could mean that participants might have to spend hours learning how to use the system to complete a single task. Thus, to meet time constraints, the participants' interfaces for a given writing task in the graph-based system would be populated with a relevant predefined template. These templates can be reproduced using the corresponding .json files for these templates, which are available in the linked GitHub repository. As a result, participants could easily drag the provided template into the interface to expand or edit it according to their needs.

\subsection{Post-study questionnaire}

Participants were asked to evaluate the two systems they had interacted with. They were asked 8 questions (4 for each system) that were rated on a 5-point Likert Scale (1: Strongly Disagree, 2: Disagree, 3: Neither Agree nor Disagree, 4: Agree, 5: Strongly Agree). Additionally, participants were asked to complete 3 short free-form questions. The Likert and free-form questions asked in the questionnaire were:

\noindent
\textbf{5-point Likert Scale Questions}
\begin{enumerate}
    \item \textbf{Satisfaction}- I was satisfied with completing the tasks using the given system.
    \item \textbf{Significance} - The interaction I had with the system significantly affected my task answer.
    \item \textbf{Ease of Use} - How would you rate the ease of using the system?  \footnote{For this question, participants were told that 1 stood for Very Easy, and 5 stood for Very Difficult. In our results, we later reverse the scale for consistency between visualisations.}
    \item \textbf{Collaboration} -I felt like I was collaborating with the system to complete the task
\end{enumerate}
\textbf{Short Answer Questions}
\begin{itemize}
    \item Compare your writing process between the two systems. For example, for either system, did you incorporate the AI generations directly into your task output? Was either system more useful for providing ideas?
    \item Was there any aspect of the graph-based system you found particularly useful while completing your task?
    \item Were there any aspects (beyond bugs or visual appearance) of the graph-based system that were frustrating? How would you address or improve these aspects? 
\end{itemize}

\subsection{Participant Demographics}
We asked for background information regarding the participant's age group, gender identity, proficiency in creative writing and familiarity with AI. For the age groups, 50\% (4/8) of the participants indicated they were 18-24, 35\% (3/8) of participants were 25-34, and 12.5\% (1/8) participants were 45-54. 75\% (6/8) of the participants identified as male, 25\% (2/8) participants identified as female. For their background in creative writing, 50\% (4/8) did not engage in creative writing activities, 37.5\% (3/8) considered themselves amateurs (write primarily for personal enjoyment or as a hobby), and 12.5\% (1/8) indicated that they were professional (have been published or paid for their writing). 87.5\% (7/8) participants had previously used AI language models like ChatGPT before, while 12.5\% (1/8) had not. 
\begin{table}[h!]
\label{table:1}
\renewcommand{\arraystretch}{1.3}
\begin{tabular}{c|p{1cm}|p{1cm}||p{1cm}|p{1cm}}
\hline
\multirow{2}{*}{} & \multicolumn{2}{c||}{\textbf{With P4}} & \multicolumn{2}{c}{\textbf{Without P4}} \\
\cline{2-5}
 & \textbf{Graph} & \textbf{Chat} & \textbf{Graph} & \textbf{Chat} \\
\hline
\textbf{Satisfaction} & 3.625 & \textbf{3.75} & \textbf{4.00} & 3.57 \\
\hline
\textbf{Significance} & 4.25 & 4.25 & \textbf{4.71} & 4.14 \\
\hline
\textbf{Ease of Use} & 2.875 & \textbf{4.125}& 3.14 & \textbf{4.00} \\
\hline
\textbf{Collaboration} & 3.375 & \textbf{3.5} & \textbf{3.71} & 3.43 \\
\hline
\end{tabular}
{\centering
\caption{Comparison of Graph-Based System vs Conversational System with and without P4}
}\end{table}

\begin{table*}[ht]
    \centering
    \begin{tabular}{c|c|l}
        \toprule
        \makecell{\textbf{Themes}} & \makecell{\textbf{Sub-Themes}} & \makecell{\textbf{Supporting Quotes}} \\ \hline
        \multirow{9}{*}{\makecell{Impact on Ideation\\and the\\ Writing Process}} 
            & \makecell{Enhanced ability \\ to conceive new \\ ideas \\ } 
            & \makecell[l]{\\[-8pt] \textbf{P1} – “The chain system was better in providing ideas as it could easily make use of \\ multiple inputs in generating a new idea.” \\ 
                          \textbf{P2} – “graph based system definitely helped me generate more ideas” \\ 
                          \textbf{P5} – “Graph based System enabled me to  be more creative and think about new ideas”} \\ \cline{2-3}

            & \makecell{Graphical\\visualisation\\provided\\direction} 
            & \makecell[l]{\\[-7pt]\textbf{P2} – “the ability to connect the nodes that helped me understand what I was thinking\\ and what direction I wanted to take” \\ 
                          \textbf{P3} – “the graph made it much easier to follow the thought process” \\ 
                          \textbf{P5}- “I liked the visualization of ideas” \\ 
                          \textbf{P6} – In response to Q2 - “Ability to create a graph based structuring of the solution”} \\ \cline{2-3}

            & \makecell{Over-\\reliance on\\ conversational\\-system} 
            & \makecell[l]{\textbf{P2} – “In the chat based system I found myself engaging less with the system and\\ asking more general questions than creative ideas.” \\ 
                          \textbf{P8} – “the conversational system was taking over the process, making me more reliant”} \\ \hline

        \multirow{7}{*}{\makecell{Challenges and \\ Limitations}} 
            & \makecell{Difficulty in \\ using Graph-\\based system} 
            & \makecell[l]{\\[-8pt]\textbf{P4} – “I was unable to use the graph system.” \\ 
                          \textbf{P7} – “Too much of cross intersection can make it very confusing.” \\ 
                          \textbf{P8} – “main issue with the custom graph based system was that it is difficult to use,\\ especially if one is using it for  the first time”} \\ \cline{2-3}

            & \makecell{Graph-based\\system's utility\\ varied} 
            & \makecell[l]{\\[-8pt]\textbf{P6} – “The conversational interface provided better and detailed answers which made \\ understanding the situation in task easy” \\ 
                          \textbf{P7} – “In my scenario the first one [chat] was more effective than the second one [graph]”} \\ \cline{2-3}

            & \makecell{UI-related \\ issues} 
            & \makecell[l]{\\[-8pt]\textbf{P6} – “It was difficult to connect nodes on horizontal scale that were far away” \\
                          \textbf{P8} – “Since I used a laptop (which has a small screen), it is difficult to make \\ the connections between the nodes without zooming out a lot”} \\ \bottomrule
    \end{tabular}
    \caption{Themes and sub-themes identified from short answers collected in survey}
    \label{tab:themes}
\end{table*}
\section{Results}
The bar plots corresponding to the questions measured on the Likert scale are depicted in Figure~2. Participants appear to be equally satisfied with their task answers for both systems. A greater proportion of participants strongly agree that the graph-based system significantly affected their answers, whereas, for the conversational system, most participants only agree. The largest difference is in the distribution of the ease of using the system. Most participants found the graph-based system much more difficult to use. On collaboration, only one participant felt that they were not collaborating with the graph-based system, whereas 3 did not feel like the process with the conversational system was collaborative.

If we take the mean of the Likert values, we get the statistics in Table~1. At first glance, the statistics indicate that our graph-based system is clearly more challenging to use and performs slightly worse than the conversational system on other metrics. However, this does not account for the fact that \textbf{P4 was unable to use the graph-based system} and thus appropriately filled in strongly disagree for all the questions related to the graph-based system. If all of P4's responses are removed (graph and conversational), we observe that the graph-based system outperforms the conversational system on every metric except ease of use. Although these differences are not statistically significant due to our small and convenience-based user sample, they provide preliminary evidence for limitations of conversational interactions.

To understand how interaction with our system differed from the conversational interface, we performed a thematic analysis of the participants' answers to the free-form short-answer questions. The identified themes, sub-themes and supporting quotes are recorded in Table~2. Our analysis shows that our system positively affected many participants' creative writing processes. This occurs because the graph-based system provides better visualisations of their interactions with the system and does not take control of the conversations participants engage in. Moreover, the system also helps participants with ideation through its unique structure. This result aligns with our expectations, as our system design was motivated by trying to expand access to the cognitive processes identified from creativity research. At the same time, we also identify various issues and limitations with our system through the analysis. Most users find the graph-based system more difficult to use, despite our attempts to keep the system simple. Some users did not find the system particularly useful. There were other minor issues with the user interface that need improvement. We further discuss our results in the discussion section.

\subsection{Example Task Answers from P3}
For illustration, we provide the task answers for P3, who completed the character description writing task. They completed the first variant with the conversational system and the second variant with the graph-based system. They included text output from the systems directly into their answers.

\subsubsection{Task answer - Character Description, Variant 1, Conversational System}
On the surface, Camilla may seem to be a normal person, with a normal job as a librarian. However, Camilla is beyond normal. Due to her photographic memory, Camilla is able to quickly analyse large amounts of data and quickly see the relationships between unrelated pieces of information. Her family has managed this historical library for generations and it has been a source of pride and stability for her family. Growing up, Camilla was surrounded by books and the sense of community that a library provides, influencing her love for literature. Her brother, however, chose a different path and became a private investigator, sharing her passion for uncovering the truth. The two often collaborated on cases, combining her research skills and his field experience to solve complex mysteries. Camilla is deemed as an over-perfectionist by her family who often worry that she obsesses too much on small details.

\subsubsection{Task answer - Character Description, Variant 2, Graph-based System}
General Kal was born into a poor family in a remote village and had to work hard from a young age to support his family. One day while working on their landlords farm, Kal witnessed his father being beaten for being unable to afford the recently increased taxes. This instilled in him a deep sense of injustice and a desire to protect his family from further oppression. Kal joined the army and with his ambition, rose through the ranks to eventually seize power. General Kal is a stern and imposing figure. His uniform is always immaculate and his eyes seem to pierce right through you, conveying a sense of authority and control. However, a closer look reveals a scar above his left eye, a reminder of a fierce battle he fought many years ago. This has left him with a slight squint in his left eye, making it difficult to see in low light conditions.

\section{Discussion}
\subsection{Ideation with the Graph-based System}
The creative writing tasks in the user study were chosen specifically because they included ideation as a key aspect of their completion. Participants reported that the graph-based system helped them generate more ideas. Some participants said this was because the visualisations of the system allowed them to better engage with their thought processes. For P2, the conversational format pre-conditioned them to ask questions instead of dealing with creative ideas. P8 mentioned that they found the ability to specify each node's randomness (temperature) essential to having more control over generating creative ideas. In contrast to these very positive experiences on ideation, P3 found both systems equally useful for ideation. Several mechanisms might affect how users use either system for ideation. We believe that while the graph-based system might demonstrate the capability to support higher levels of ideation, exploiting these is much more dependent on the user. For example, the user can set up the same template with different inputs and see how the output varies. This isn't possible with a conversational interface. 

\subsection{Visualising the Creative Writing process}
Beyond the graph-based system's ability to generate ideas and text, participants commented on how the graph-based system provided a visualisation that offered direction and guidance. One of the main contributions of the AI-Chains \cite{aichains} paper was that breaking down problems into well-defined chaining steps provided guardrails that kept users on track. These guardrails also prevented free exploration.

Because of our flexible node definitions, our graph-based system provided direction and guidance to the user without restricting free exploration. Additionally, for many participants, the graph-based structure also acted as a tool for visual organisation. The graphical representation of thoughts and ideas as nodes allowed some users (specifically, P2, P3, P5, P6 with quotes in Table~2) to better complete their tasks. This ability to organise thoughts and ideas also means a higher learning curve to use the system. We believe this higher learning curve is responsible for many of the difficulties other participants faced while using our system. We discuss these in the limitations section.

\section{Limitations}
The most significant limitation of our study is the limited number of participants (8) in our user study. While this might render some of our quantitative results insignificant, we still believe this paper's qualitative analyses can guide future user interfaces that seek to promote ideation with LLMs. Another key limitation of our system is that it is more difficult to use. Conversational interfaces are much more prevalent in everyday life than graph-node interfaces. Even if our specific graph-based system requires little additional knowledge,  when compared to a conversational interface, navigating the system can prove difficult. P4 was completely unable to engage with the graph-based system. In their background, they indicated that they had not used any AI tools before. Yet, they could easily complete the creative writing task with the conversational interface. On the other hand, P3 stated, "There were no aspects of the graph-based system that I found frustrating, it was generally easy and intuitive to use." The difficulty of the graph-based system might depend on the participant's past experience with graph-node interfaces. Some level of difficulty was also introduced by the imperfect user interface. Some participants, such as P6 and P7, found the conversational interface to be better. We suspect this could be a direct consequence of the ease of using the system. It is also possible that conversational and graph-based systems appeal to different categories of users, and regardless of any changes that can be made, there will always be some participants who prefer one over the other. Moreover, this project used a relatively small LLM due to limited compute availability. As most users were still able to generate and write creative stories, we believe this limitation is not that restricting. Finally, our participants were largely either amateur creative writers or had no creative writing experience. LLMs have been shown to affect the individual creativity of amateur writers more effectively than professional writers  \cite{indi}. To this extent, the results of our system do not generalise to professional creative writers and only apply to hobbyists. 


\section{Future work}
The types of pre-defined nodes in the current system can be further expanded to explore different use cases in creative fields beyond creative writing. For example, incorporating image-to-text and image-to-image models could allow image-based input to be added to chains. This would allow chains to have more significant applications in creative fields where images are the primary medium. 

Since the start of this project, multimodal models that integrate multiple forms of media have gained prevalence. It would be intriguing to explore how multimodality could play a role in such an interface. Our research utilised smaller 7B parameter LLMs. Exploring the impact of larger LLMs on chains, given their extensive context length, is also valuable. With the increasing accuracy of the larger LLMs, maintaining global context while still being able to map out graphs for a writer's thought process might be a more reasonable approach instead of each node maintaining individual context.

The flexibility offered by the interface would also be interesting to experiment with. When given the capability to switch between different LLMs and text-to-image models for each node, can creative writers leverage specialised and fine-tuned models for enhanced controllability?

\section{Conclusion}
In conclusion, we developed a novel, graphical, node-based system for chaining generative AI models. This system was designed to overcome the structural constraints of traditional chat-based interactions. Creativity research and previous research on chaining were used as a basis for our design. We demonstrated the system's potential through a user study, showcasing how it can assist in ideation and visualisation to enhance the creative writing process. The results were discussed, highlighting the meaning they possess for research in future interfaces that seek to promote ideation. While there are several limitations to our system and user study, our research opens up new possibilities for ideation and interaction with generative AI systems.

\section{Author Contributions}
Author 1 (AS) designed the system and drafted the paper. Authors 2 (MTL) and 3 (JM) supervised the research and provided guidance and direction for the development of this work. 







\bibliographystyle{iccc}
\bibliography{iccc}

\end{document}